# Software Engineering Meets Network Engineering: Conceptual Model for Events Monitoring and Logging

**Sabah Al-Fedaghi and Bader Behbehani**
*sabah.alfedaghi@ku.edu.kw　　bader92@gmail.com*
Computer Engineering Department, Kuwait University, Kuwait

**Summary**
Abstraction applied in computer networking hides network details behind a well-defined representation by building a model that captures an essential aspect of the network system. Two current methods of representation are available, one based on graph theory, where a network node is reduced to a point in a graph, and the other the use of non-methodological iconic depictions such as human heads, walls, towers or computer racks. In this paper, we adopt an abstract representation methodology, the thinging machine (TM), proposed in software engineering to model computer networks. TM defines a single coherent network architecture and topology that is constituted from only five generic actions with two types of arrows. Without loss of generality, this paper applies TM to model the area of network monitoring in packet-mode transmission. Complex network documents are difficult to maintain and are not guaranteed to mirror actual situations. Network monitoring is constant monitoring for and alerting of malfunctions, failures, stoppages or suspicious activities in a network system. Current monitoring systems are built on ad hoc descriptions that lack systemization. The TM model of monitoring presents a theoretical foundation integrated with events and behavior descriptions. To investigate TM modeling's feasibility, we apply it to an existing computer network in a Kuwaiti enterprise to create an integrated network system that includes hardware, software and communication facilities. The final specifications point to TM modeling's viability in the computer networking field.

## 1. Introduction

This paper is a continuation of a previous article [1]. Our previous paper dealt with network documentation. This paper presents research in the area of networking monitoring in packet-mode transmission using the same conceptual modeling method, called a thinging machine (TM).

Current depictions of network diagrams have been developed over many years and include hundreds of various symbols, which range from walls to computer screens to server racks to a cloud-based storage system. Network diagrams may also be based on an abstract graph theory representation, which presents a network as a set of nodes and edges. Scholars have been greatly interested in defining a single coherent representation of a network [2]. This effort requires a rich understanding of relationships among network elements according to their applicability, environment and internal functionalities.

1.1 Monitoring Computer Networks

In general, "monitoring" refers to the methodical and continuing gathering, examination and use of information for management control and decision-making processes [3]. In computer networks, continuous management requires monitoring the network through apparatuses that permit supervisors to instantaneously access information on the system's state and handle alerts for events such as malfunctions, failures, stoppages or suspicious activities [4]. A network monitoring system monitors external security threats and internal incidents caused by system crashes and malfunctions of servers, connections or other devices.

From a technical perspective, the world is becoming increasingly interconnected. Managing networks with such interconnectivity is an essential factor for maintainability and sustainability. Network maintenance requires a rapid pinpointing process as well as addressing any difficulty regardless of whether a failing mail daemon or a broken fiber optic connection caused the incident. A modern network system offers information about its events. These events are generated by

- Logs within operating systems
- Archiving records of events within servers
- Logs of errors, warnings and failures within applications
- Records of suspicious traffic generated by firewalls and virtual private-network gateways
- Observing traffic among various network segments in network routers and switches
- Simple network management protocol (SNMP) traps and alerts reported to a management console [5].

Events occurring within a network's components are recorded as logs related to network security. These events are created by various sources such as applications, networking equipment, antivirus software, firewalls and intrusion detection systems, operating systems and workstations [6]. Log management is critical when creating and handling adequately detailed event data in a suitable period of time. Log analysis is valuable for isolating security instances, detecting policy abuses, capturing suspicious action and identifying working difficulties. Logs are also precious when inspecting, investigating and recognizing long-term problems.

The involved network devices monitor their own behavior, receive and relay messages from other networks and duplicate alerts. Event logs register all the activities on the system's devices for the purpose of utilizing the generated data to address security and performance issues. The result is a complicated picture of events and behaviors such that a single problem can produce a blizzard of alerts and event messages. In this situation, a huge amount of data is created all over the network that no one can easily scan and handle alone. According to Kay [5], in 2000, a Computer Sciences Corp OC-12 (a fiber optic connection) could generate about 850 MB of event data in an hour, which translates into more than 600 GB of data per month. In 2020, the volume of data being created generally increased with the introduction of 5G networks, edge computing, the Internet of Things and the increased adoption of cloud computing. IDC expected that 59 ZB of data would be generated and handled in 2020, up from 41 ZB in 2019 [7].

Currently, the same iconic representations used to describe computer networks are also used to represent monitoring systems. For example, the network monitoring server OpManager is described in textual language, and its architectural structure is given as shown in Fig. 1 [8]. Taking another example, Leskiw [9] described Syslog as a mechanism that allows network devices to communicate with a logging server (Syslog server) by sending event messages. Various event message types are logged using Syslog protocol, which covers a wide range of network devices. Most network equipment, such as routers and switches, can send Syslog messages [9]. Syslog simply sends messages to a central location when specific events are triggered. Syslog processes are represented as shown in Fig. 2.

Network management also relies on an enhanced, coherent high-level model for monitoring of computer networks. High-level model architecture should demonstrate the event-based mechanism within a monitoring system. Events are produced by capturing records of data generated by network components such as logs of traffic, malfunctions, failures, system faults, breaches and network address translations (NATs). The visualization of logs' sources and details that network components produce in a monitoring system is completely vague. Network management lacks a high-level model to represent a monitoring system for network components. Network administrators and decision makers should be able to visualize events within the monitoring system using a high-level modeling approach that helps them understand points of failure, business weaknesses and opportunities for enhancement.

1.2 Proposed Thinging Machine Approach

TM adopts a high-level representation methodology to model computer networks. It defines a single coherent network architecture and topology similar to engineering schematics. TM topology consists of only five generic actions with two types of flow arrows. The TM model of monitoring presents a theoretical foundation that is integrated with a network's events and behavior descriptions.

In this paper, to investigate TM modeling's feasibility, we applied TM-based modeling to an existing computer network in a Kuwaiti enterprise to create an integrated network system that includes hardware, software and communication facilities.

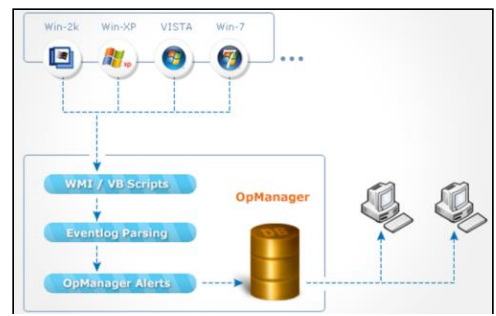

Fig. 1 Sample network monitoring description (From [8]).

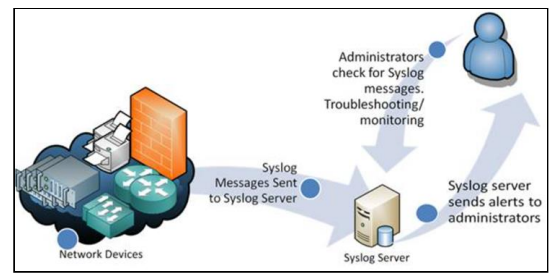

Fig. 2 Another representation of a monitoring system (From [9]).

Specifically, in this paper, we focus on modeling packet flows through various software and devices in the network. Accordingly, the monitoring problem is limited to monitoring packets in the network instrument. For monitoring purposes, we propose a more fundamental approach for high-level description of computer networks. Our model involves conceptualizing the structure of network diagrams, and the resultant schema is used as a vehicle for communication among engineers, managers and decision makers.

1.3 Outline

- In the next section, we summarize TM [10].
- Next, we discuss a case study involving network monitoring. The case involves an IT department that has acquired a large number of servers and more storage capacity, creating a well-defined and understandable network description, as discussed in our previous paper [10]. Due to space limitations, we concentrate on a specific part that is closely related to monitoring in its functionality.
- We focus on an adaptive security appliance (ASA) network, including a flowchart of the ASA packet process algorithm. Specifically, we concentrate on modeling the FirePOWER services module.
- Therefore, our main contribution in this paper concerns monitoring in the FirePOWER service module. We use TM to model FirePOWER as a part of the network and provide a general description of the system's context, with further focus on the LINA subsystem of the FirePOWER system.

## 2. Thinging Machine Modeling

Diagrammatic modeling languages hold great promise for software and network engineering and can depict structural and behavioral specifications. However, many practitioners still consider diagrammatic languages mere "doodles" [11].

TM uses a diagramming language built on one-category ontology and five actions that, when designated over time, produce a behavioral model in terms of the chronology of events. The TM model articulates the ontology of the world in terms of an entity called a *thimac* (the first three letters of *thing* and *machine*), with double faces or two "being-nesses," as a *thing* and simultaneously as a *machine*. The first side of the coin exhibits the wholeness (thing) assumed by the thimac; on the second side, operational symptoms (processes) emerge, providing a "force" that goes beyond structures or things to embrace other things in the thimac. To conceptualize a thimac as a thing presents no indication as to the content of the thing whereas to conceptualize a thimac as a machine forces a definite structure of actions with a flow of other things (See Fig. 3).

A thing is subjected to doing (e.g., a tree is a thing being planted or cut), and a machine does (e.g., a tree is a machine that absorbs carbon dioxide and uses sunlight to make oxygen). The thing tree and the machine tree are two faces of the thimac tree. A thing is viewed based on Heidegger's [12] notion of thinging. A thing is a machine, and a machine is a thing. Fig. 4 shows a complete generic machine. The actions in the machine (also called stages) can be described as follows:

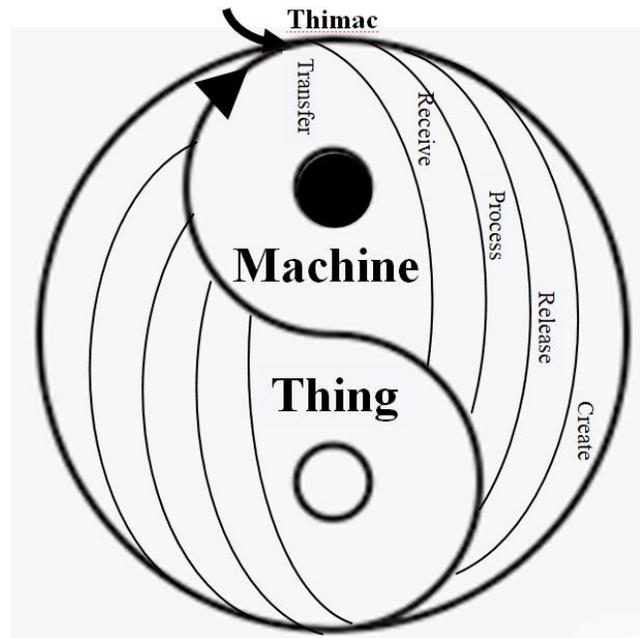

Fig. 3. Thimac as a yin-yang symbol. A thing and a machine are each a version of the other.

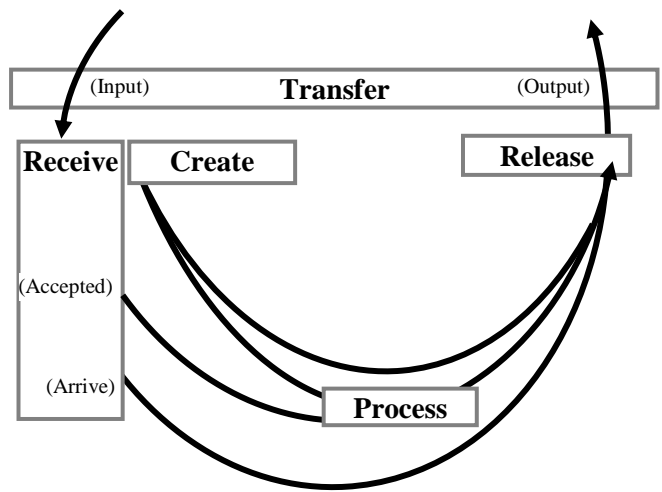

Fig. 4 Flow of things in a thinging-machine model.

**Arrive:** A thing moves to a new machine.
**Accept**: A thing enters a machine. For simplification, we assume that all arriving things are accepted; therefore, we can combine, arrive and accept the thing as the **Receive** stage.
**Release**: A thing is marked as ready to be transferred outside the machine (e.g., in an airport, passengers wait to board after passport clearance).
**Process**: A thing is changed in form, but no new thing results.
**Create**: A new thing is born in a machine.
**Transfer**: A thing is input into or output from a machine.

Additionally, the TM model includes storage and triggering (denoted by a dashed arrow in this paper's figures), which initiates a flow from one machine to another. Multiple machines can interact with each other through movement of things or triggering stages. Triggering is a transformation from one series of movements to another (e.g., electricity triggers cold air).

## 3. Thinging Machine-Based Model of Network Monitoring

In this section, we discuss a case study involving network monitoring. Demand for a live network-monitoring behavioral system is growing rapidly among network administrators. Administrators require appropriate, efficient and accurate logging systems with the ability to maintain, upgrade and manage all network resources. This section includes several discussions introducing the subsystem that we will model. Additionally, we discuss the research area that motivated our proposed model. Moreover, we demonstrate TM modeling for static, dynamic, behavioral and monitoring representations of the subsystem.

### 3.1 Background of the Modeled System

In this section, we focus on a case study of a single organization in Kuwait. The organization's business requirements are growing, causing an increase in service demands for information technology (IT) resources. The IT department has acquired a large number of servers and greater storage capacity. Creating a well-defined and understandable network description is a way to visualize an organization's network structure and is critical for improving the efficiency, effectiveness and timeliness of maintenance activities. This need for improvement motivates the development of a TM-based model instead of the use of symbols such as walls, towers and human and computer icons, which do not produce systematic depictions that define coherent network architecture.

Additionally, using abstract network architecture diagrams (graphs) of nodes and lines—which focus only on presenting the communication between the nodes—is equally unsatisfactory: their extremely abstract content does not expose the nodes' internal functionalities. The individual features of the nodes' static and dynamic aspects are totally absent from a graph. In this case, a TM can form the foundation of a general description of topological connectivity within the network, which can be utilized to understand the network, communication among various types of stakeholders, the maintenance process, monitoring and documentation.

In pursuing this goal, we model packet flows through various sub-networks as a single phenomenon that ties various components of the network system together. TM modeling represents the network as a thimac. We model the network as an existing system. Upon inspecting the current network, we divided the TM thimac into several sub-thimacs. Al-Fedaghi and Behbehani discussed the current network (Fig. 5) in detail in [1].

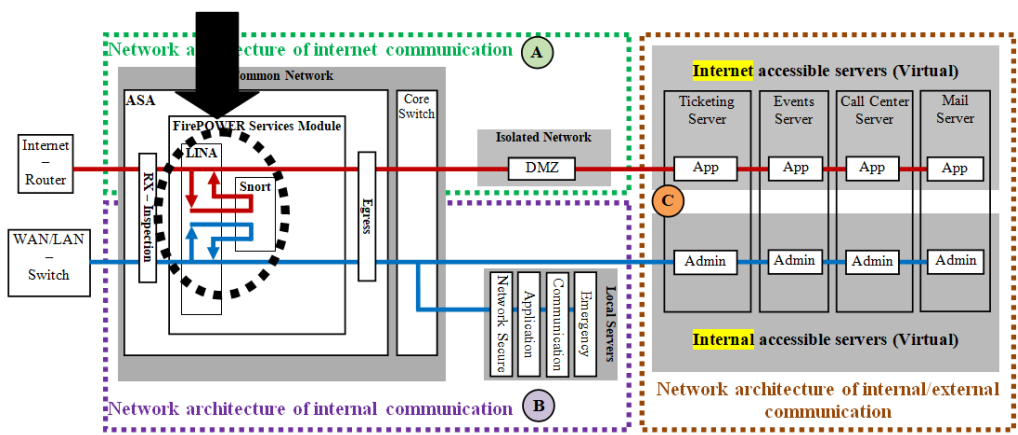

Fig. 5 General description of case study network architecture.

In this paper, we focus on adaptive security appliance (ASA). ASA is a basic component used in many systems that allows users to securely access data and other resources within the network while delivering enterprise firewall capabilities. The ASA 5500-X manual (2015) includes a flowchart of the ASA packet process algorithm, as shown in Fig. 6 and partially in Fig. 5. Al-Fedaghi and Behbehani model this algorithm in TM (2020), except for the FirePOWER services module shown in blue in Fig. 6.

3.2 Monitoring in FirePOWER Service Module

This section includes our main contribution in this paper. We use TM to model FirePOWER as a part of the network. We provide a general description of the context of the FirePOWER system, with further focus on the LINA subsystem (Fig. 5). Accordingly, in this section, we present

1. Static TM model: Al-Fedaghi and Behbehani (2020) presented this type of modeling to specify ASA and other components in Fig. 5. In this paper, we focus on the TM static model of the FirePOWER services module.
2. Events/dynamic model: This level of modeling specifies the events of packets' journey through LINA.
3. Behavior model: This level of modeling specifies the chronology of events identified in (2).

3.3 General Description of the FirePOWER System

ASA aggregates security capabilities into one device (see Fig. 7). These capabilities include firewall, antivirus, intrusion prevention and VPN capabilities.

ASA's purpose is to stop security threats and attacks within the network. The FirePOWER service helps discover vulnerabilities before an attack takes place by detecting, blocking and defending against network security attacks. It provides the following key capabilities: access control, intrusion detection and prevention, advanced malware protection (AMP) and file control. Ingress packets are processed against access control lists (ACLs), connection tables, network address translation (NAT) and application inspections before traffic is forwarded to the FirePOWER services module.

The so-called Firepower Threat Defense (FTD) is the main software system that actively runs the FirePOWER service in FirePOWER (Fig. 7). Some of the FirePOWER functions found in FTD are the intrusion prevention system (IPS), application visibility and control (AVC), URL filtering and AMP. These services are called the "FirePOWER services module" in Fig. 7. Fig. 8 shows a sample description from [14].

FTD software consists of two main engines, along which packets will flow. The first is the LINA engine (See Fig. 9).

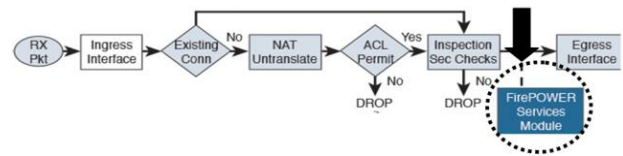

Fig. 6 ASA packet process algorithm (adapted from [13]).

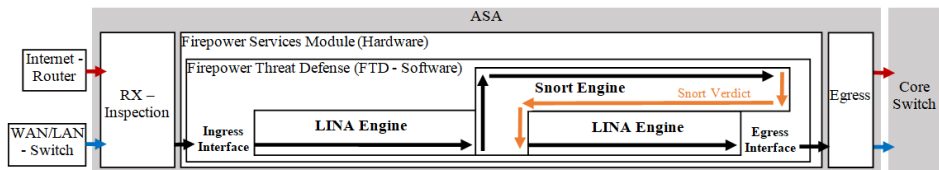

Fig. 8. Sample description of FTD packetprocess [14]

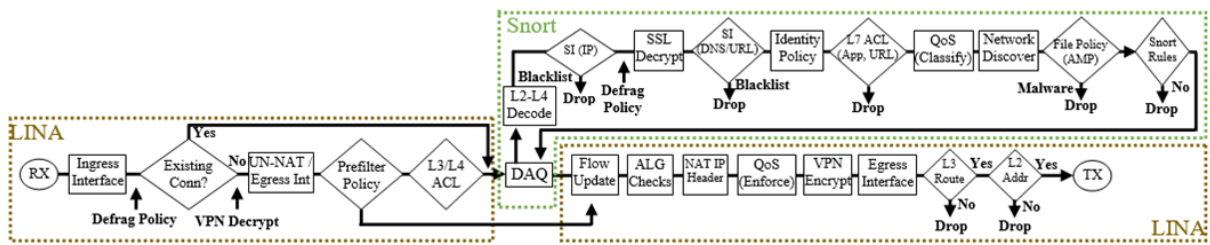

Fig. 7 ASA with FirePOWER packet process.

Fig. 9 Detailed FTD packet process.

which handles the packet's entry routing (see Fig. 10 for simplified packet structure) by controlling the outer IP header via the inspecting traffic tunnel. The LINA engine inspects the traffic tunnel between systems by checking the packet's outer header (see relevant details of packet structure in Fig. 10) and deciding whether incoming packets are allowed access. The second engine is the Snort engine, in which packets can reach the LINA engine's allowance permissions. The Snort engine handles packet entry routing by inspecting the inner IP header.

Prefilter policies handle outer IP header policies within the LINA engine, and access control policies handle inner IP header policies within the Snort engine. For instance, the type of policies applied to traffic tunnels (GRE, IP-in-IP, etc.) are prefilter policies; policies applied inside the sessions are access control policies.

## 4. Modeling the LINA Subsystem in a Thinging Machine

One of the main engines in FTD software is LINA, which is responsible for handling the execution of trusted connections, NAT, prefilter policies, etc. LINA is also responsible for deciding whether the packet should be further processed in Snort, depending on the predefined rules and policies set within its components [15]. This section presents LINA's static, dynamic, behavioral and monitoring models.

4.1 Static Model

Fig. 11 shows the static TM model of LINA. In the figure, the packet flows from the Internet or WAN/LAN (Circle 1), reaching ASA, where the packet is processed starting from RX to inspection checks (see Fig. 6). Then, the packet flows to LINA (2) and to the ingress interface (3), which handles packet entry. The packet is processed (4) as follows.

*Packet Entry*
A. The input counter for incoming packets is triggered (5) and incremented by one (6).
B. The destination is extracted (9) and transferred (10) for comparison with LINA's list of destinations (11).

This comparison involves the packet's destination (12), with one destination fetched from the list (13) at a time. The comparison process (14) involves the following:
- If destinations are different and not at the end of the list (the list contains more destination entries), the next destination in the list is fetched for a new comparison (15).
- If the destinations are different and at the end of the list (16),
    a. The packet is processed to be decrypted (17) and then sent to the Untranslate–NAT module (18).
    b. Defragmentation of payload: The stored (8) payload (data) is released (19) to flow to the defragmentation module (20), where it is defragmented (21), removing spaces, and the new payload (22) flows (23) to be stored (8).
- If the destinations are the same (24),
    a. The defragmentation of the payload (19 to 23) is performed.
    b. The packet is released (25) to flow (26) to DAQ (Data Acquisition).

*Untranslate Network Address Translation*
The packet is received (27) and processed (28) to extract its header (29). The header is processed (30) to extract its destination (31), which flows (32) to be compared with destinations in a NAT table. The NAT table is processed (33), and a destination address is retrieved (34) and transferred (35) for comparison (36).

- If the destination is not in the table (37) and is not at the end of the table, a new address is retrieved from the table to be compared with the destination.
- If the destination is not in the table (38) or at the end of the table, the egress interface (39; discard Untranslate NAT process) is skipped and the incoming destination packet flows to the prefilter policy module (see below).
- If the destination exists in the address, the process will result in YES (40), which will release and transfer the destination to the egress interface.

In the egress interface (39), a table of global routes (41) exists to show each destination with its related route. The global route table is processed (42) to retrieve (43) one route that flows (44) for comparison. The incoming destination (40) and the route (44) are compared (45).

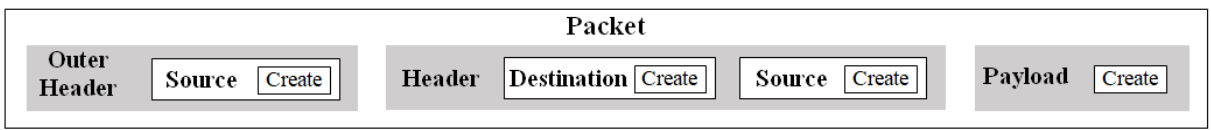

Fig. 10 Packet content upon entering ASA.

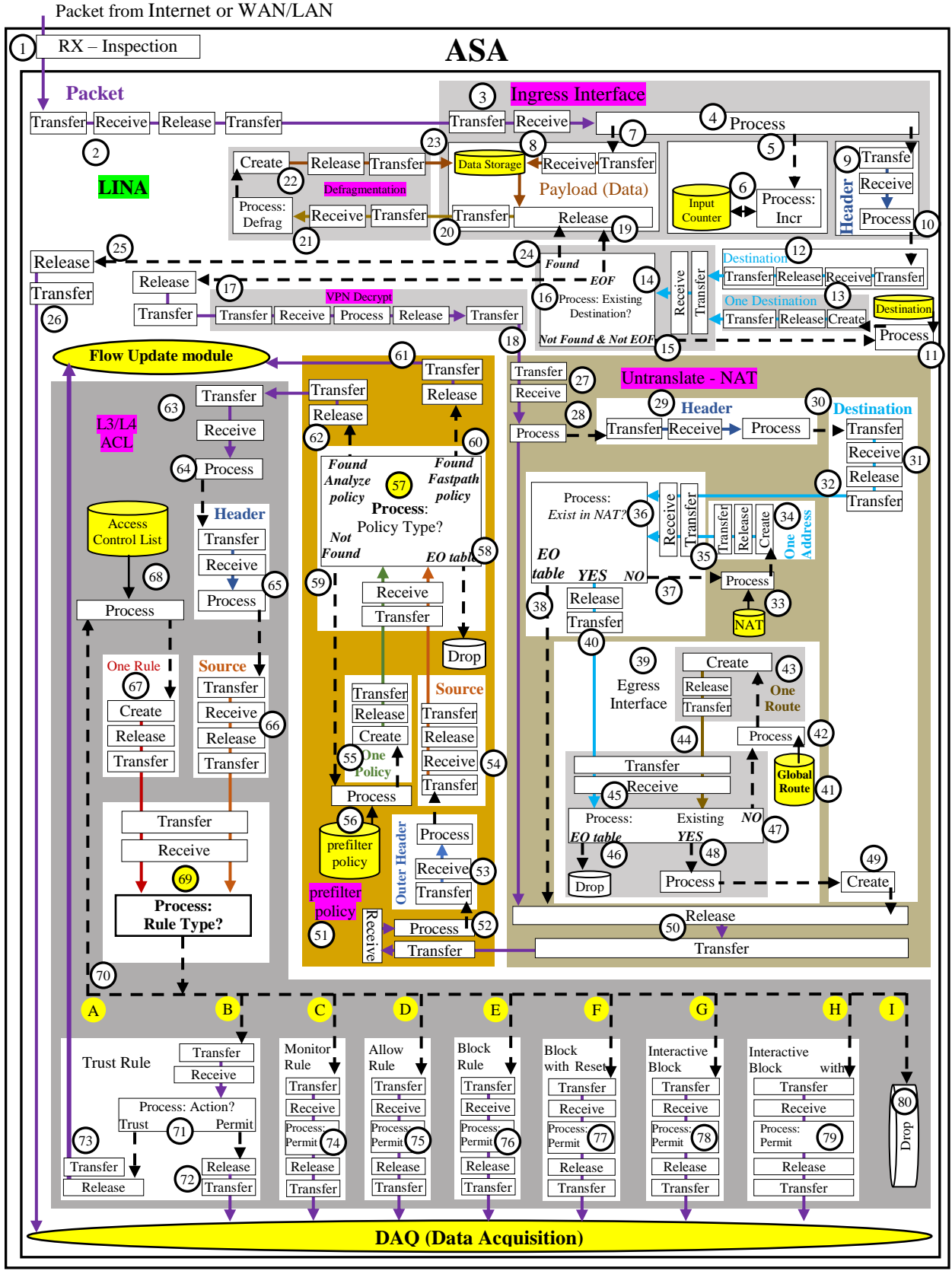

Fig. 11 FTD static TM.

- If the destination does not exist in the route and the global route table contains no remaining routes (46), the incoming destination is dropped.
- If the destination does not exist in the route and the global route table contains more routes (47), another route is retrieved from the table to be compared with the incoming destination.
- If the destination exists in the route (48), then a trigger to create (49) a new destination is performed based on the route. The packet with the new destination is released (50) to the prefilter policy module (51).

*Prefilter Policy (Traffic Filtering)*

The packet in the prefilter policy module is processed (52) to extract the outer header (53), which in turn is used to extract the source (54). The source and the policy (55) that are retrieved from the prefilter policy table (56) are compared (57).

- If the source is not included in the policy and the prefilter policy table contains no remaining policies, the involved packet is dropped (58).
- If the source is not included in the policy and the prefilter policies table contains more policies, a new policy is retrieved to be compared with the source (59).
- If the source is included as a "fastpath" policy (60 – a policy is found), the packet is released to the flow update module (61).
- If the source is included as an "analyze" policy (62 – a policy is found) the packet is released to the L3/L4 ACL (63) module.

*L3/L4 Access Control List*

The packet in the L3/L4 ACL module is processed (64) to extract the header (65), which in turn is used to extract the source (66). The source and the access control rule (67) that are retrieved from the ACL (i.e., trust, monitor, allow, block, block with reset, interactive block or interactive block with reset) (68) are compared (69).

A. **No type**: If no rule is applicable to the source and the ACL still has rules to be examined, a new rule is retrieved to be compared with the source (70).
B. **Trust rule:** If the trust rule is applicable to the source, the involved packet is processed (71) as follows.
   - If trusted, the packet is released (73) to a **flow update module** (another diagram in modeling the system).
   - If permitted, the packet is released (72) to **DAQ** (another diagram in modeling the system).
C. **Monitor rule:** If the monitor rule is applicable to the source, the involved packet is processed (74) and sent to **DAQ**.
D. **Allow rule:** If the allow rule is applicable to the source, the involved packet is processed (75) and flows to **DAQ**.
E. **Block rule:** If the block rule is applicable to the source, the involved packet is processed (76) and flows to **DAQ**.
F. **Block with Reset rule:** If the block with reset rule is applicable to the source, the involved packet (including a reset rule) is processed (77) and flows to **DAQ**.
G. **Interactive Block rule:** If the interactive block rule is applicable to the source, the involved packet (including a bypass rule) is processed (78) and flows to **DAQ**.
H. **Interactive Block with Reset rule:** If the interactive block with reset rule is applicable to the source, the involved packet (includes an intersect rule) is processed (79) and flows to **DAQ**.
I. **Deny:** If any rule is applicable to the source under a deny action, the involved packed is dropped (80).

4.2 Dynamic Model

In this model, we select a decomposition of the static model to identify the events embedded in the description. The decompositions chosen are as follows (See Fig. 12):

Event 1 ($E_1$): A packet's arrival to LINA.
Event 2 ($E_2$): The packet flows to the ingress interface.
Event 3 ($E_3$): The packet's details are processed.
Event 4 ($E_4$): The ingress interface's input counter is incremented.
Event 5 ($E_5$): The payload is extracted and stored.
Event 6 ($E_6$): The header is extracted.
Event 7 ($E_7$): The destination is extracted.
Event 8 ($E_8$): The destination flows to be compared.
Event 9 ($E_9$): A destination form is retrieved from the destination table.
Event 10 ($E_{10}$): The retrieved destination flows to be compared.
Event 11 ($E_{11}$): The incoming destination and the retrieved destination are compared.
Event 12 ($E_{12}$): A new destination is retrieved from the destination table.
Event 13 ($E_{13}$): The incoming destination does not exist in the destination table.
Event 14 ($E_{14}$): The incoming destination is found in the destination table.
Event 15 ($E_{15}$): The payload is retrieved from data storage and flows to defragmentation.
Event 16 ($E_{16}$): The payload is defragmented and stored.
Event 17 ($E_{17}$): The packet flows to DAQ.
Event 18 ($E_{18}$): The packet flows to VPN decrypt.

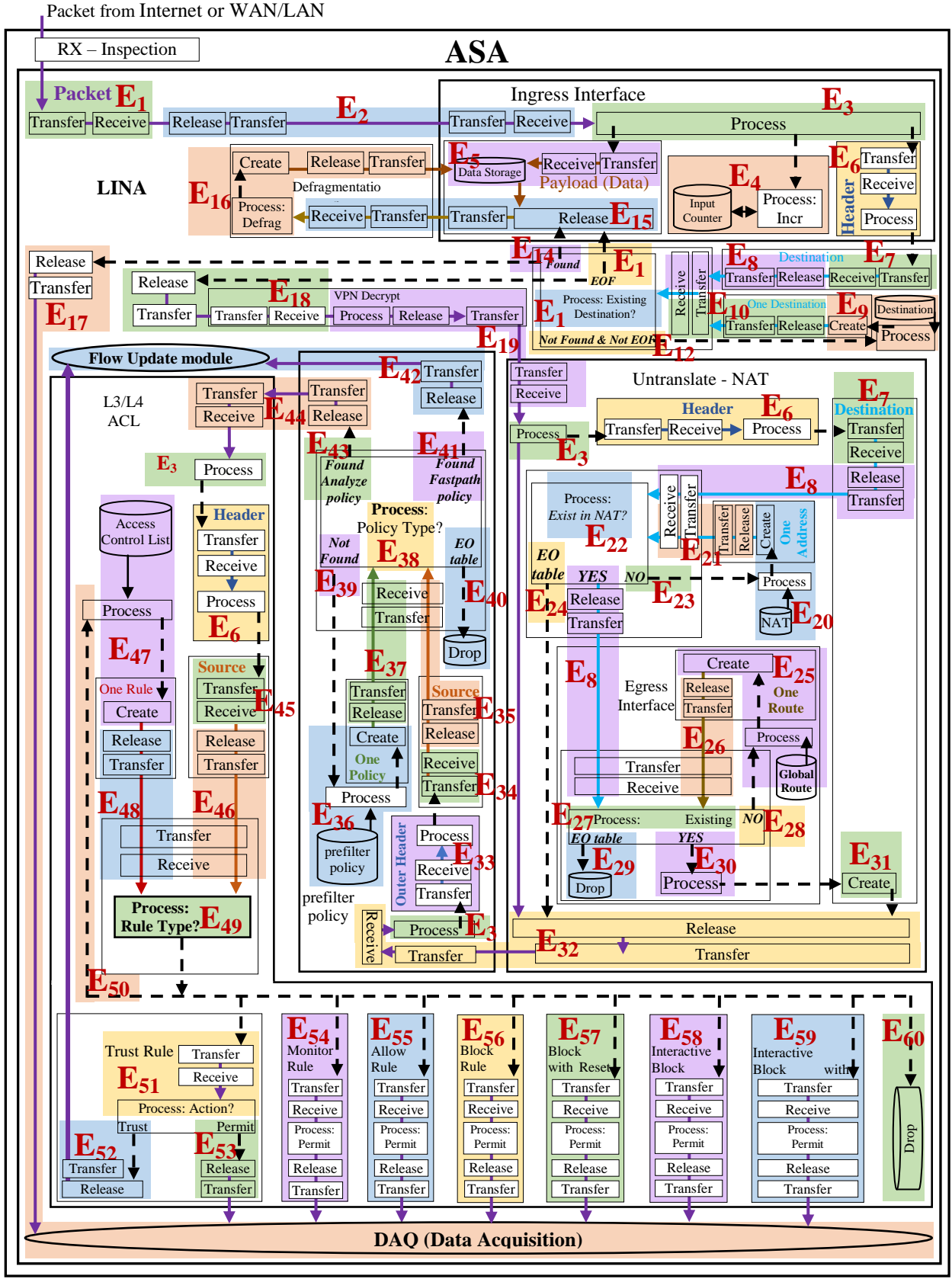

Fig. 12 FTD event model.

Event 19 ($E_{19}$): The packet is decrypted and flows to Untranslate–NAT.
Event 20 ($E_{20}$): A destination address is retrieved from the NAT table.
Event 21 ($E_{21}$): The retrieved destination address flows to be compared.
Event 22 ($E_{22}$): The incoming destination and the retrieved destination address are compared.
Event 23 ($E_{23}$): A new destination address is retrieved from the NAT table.
Event 24 ($E_{24}$): The incoming destination does not exist in the NAT table.
Event 25 ($E_{25}$): A route is retrieved from the global route table.
Event 26 ($E_{26}$): The retrieved route flows to be compared.
Event 27 ($E_{27}$): The incoming destination and the retrieved route are compared.
Event 28 ($E_{28}$): A new route is retrieved from the global route table.
Event 29 ($E_{29}$): The incoming destination is not included in the global route table, and the packet is dropped.
Event 30 ($E_{30}$): The incoming destination is found in the global route table.
Event 31 ($E_{31}$): A new destination is set for the incoming packet.
Event 32 ($E_{32}$): The packet flows to the prefilter policy.
Event 33 ($E_{33}$): The outer header is extracted.
Event 34 ($E_{34}$): The source is extracted.
Event 35 ($E_{35}$): The source flows to be compared.
Event 36 ($E_{36}$): A policy is retrieved from the prefilter policy table.
Event 37 ($E_{37}$): The retrieved policy flows to be compared.
Event 38 ($E_{38}$): The source and the policy are compared.
Event 39 ($E_{39}$): A new policy is retrieved from the prefilter policy table.
Event 40 ($E_{40}$): The source is not included in the prefilter policy table, and the packet is dropped.
Event 41 ($E_{41}$): The source is found in the prefilter policy table with a fastpath policy.
Event 42 ($E_{42}$): The packet flows to the flow update module.
Event 43 ($E_{43}$): The source is found in the prefilter policy table with an analyze policy.
Event 44 ($E_{44}$): The packet flows to L3/L4 ACL.
Event 45 ($E_{45}$): The source is extracted.
Event 46 ($E_{46}$): The source flows to be compared.
Event 47 ($E_{47}$): A rule is retrieved from ACL.
Event 48 ($E_{48}$): The retrieved rule flows to be compared.
Event 49 ($E_{49}$): The source and the rule are compared.
Event 50 ($E_{50}$): A new rule is retrieved from ACL.
Event 51 ($E_{51}$): The source is found in ACL with a trust rule, and an action is performed on the packet.
Event 52 ($E_{52}$): The packet flows to the flow update module using a trust action.
Event 53 ($E_{53}$): The packet flows to DAQ using a permit action.
Event 54 ($E_{54}$): The source is found in ACL with a monitor rule, and the packet flows to DAQ using a permit action.
Event 55 ($E_{55}$): The source is found in ACL with an allow rule, and the packet flows to DAQ using a permit action.
Event 56 ($E_{56}$): The source is found in ACL with a block rule, and the packet flows to DAQ using a permit action.
Event 57 ($E_{57}$): The source is found in ACL with a block with reset rule, and the packet flows to DAQ using a permit action.
Event 58 ($E_{58}$): The source is found in ACL with an interactive block rule, and the packet flows to DAQ using a permit action.
Event 59 ($E_{59}$): The source is found in ACL with an interactive block with reset rule, and the packet flows to DAQ using a permit action.
Event 60 ($E_{60}$): The source is found in ACL with a deny action, and the packet is dropped.

4.3 Behavioral Model

Fig. 13 shows the behavioral model based on the decompositions in the dynamic model and according to the chronology of events.

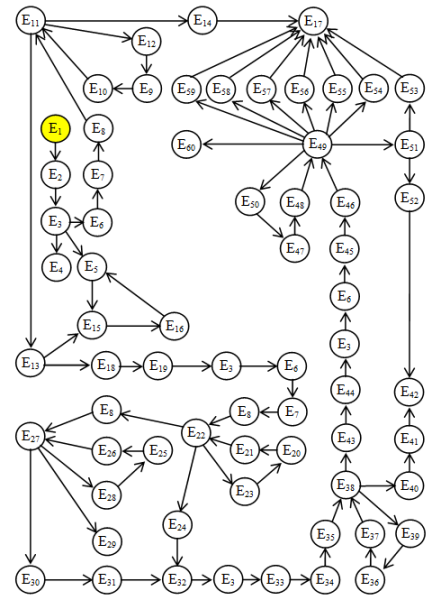

Fig. 13 TM behavioral model

4.4 Monitoring Model

Monitoring can be applied to all events, $E_1, E_2, \ldots E_{44}$, or a subset of these events (see Fig. 14). Accordingly, when the event $E_i$ happens, it triggers a meta-event (an event that is caused by an event), signified as $M_i$. For instance, $M_9$ creates a record of $E_9$ that contains data about the time of $E_9$ and other data—changes in values, alerts, warnings or any other needed information. The time data can contain various time information (e.g., start/end time, period; Fig. 15). A log manager may contain all sets of meta-events to create temporal log registrations for historic archives of all events, or it can merge events into one bigger event. For example, a history record can be generated for the packet in the ingress region ($E_3, \ldots E_6$).

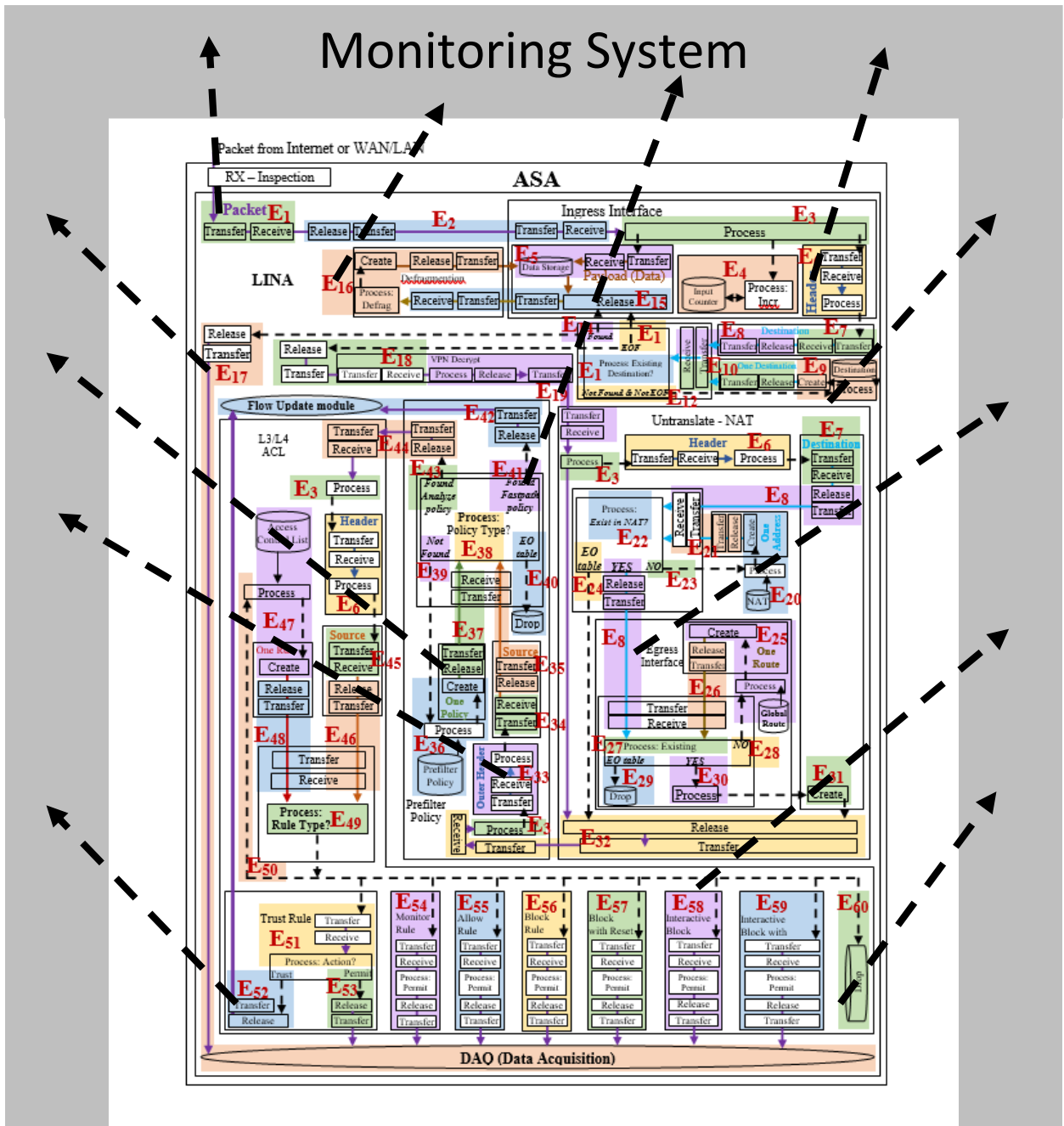

Fig. 14. General overview of the monitoring system

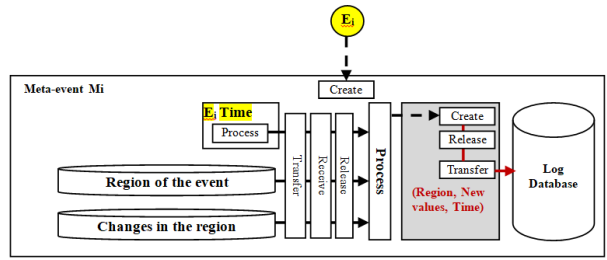

Fig. 15 Generating temporal data for changes in balance values.

## CONCLUSION

We applied TM modeling to monitoring in packet-mode transmission. Continuity is required in networking, especially for alert processes upon failures, stoppages or suspicious activities within a network system. Currently, monitoring systems lack the conceptual representation and systemization that generate proper event logs that can precisely describe internal communications within network resources. We applied TM-based modeling to an existing computer network in an enterprise in Kuwait to create an integrated network system including hardware, software and communication facilities. The results speak for themselves: we can apply a single modeling methodology with a simple ontology of five actions and two types of arrows uniformly to all stages of static, dynamic, behavioral and monitoring representations. Of course, TM is still a theoretical artifact that needs to be implemented in reality. Further research in this direction will involve building computer-based tools that can facilitate building such a model.

## References

[1] Al-Fedaghi, S., Behbehani, B.: *How to Document Computer Networks*. Journal of Computer Science 16(6), 723–734 (2020). DOI:10.3844/jcssp.2020.723.434

[2] Wolf, T., Griffioen, J., Calvert, K. L., Dutta, R., Rouskas, G. N., Baldine, I., Nagurney, A.: *Choice as a Principle in Network Architecture*. In: Proc. of the ACM SIGCOMM 2012 Conference on Applications, Technologies, Architectures, and Protocols for Computer Communication (2012)

[3] Umhlaba Development Services: *Introduction to Monitoring and Evaluation Using the Logical Framework Approach.* Noswal Hall, Braamfontein, Johannesburg, South Africa (2017) https://eeas.europa.eu/archives/delegations/ethiopia/documents/eu_ethiopia/ressources/m_e_manual_en.pdf

[4] Svoboda, J., Ghafir, I., Prenosil, V.: *Network Monitoring Approaches: An Overview*. Int J Adv Comput Netw Secur 5(2), 88–93 (2015). DOI: 10.15224/978-1-63248-061-3-72

[5] Kay, R. *Event Correlation*. In: Computerworld (2003). https://www.computerworld.com/article/2572180/event-correlation.html

[6] Kent, K., Souppaya, M.: *Guide to Computer Security Log Management*. NIST special publication 92, 1–72 (2006)

[7] O'Brien, C.: *5 IPOs That Show the Importance of Data in 2020.* In: VentureBeat (2020). https://venturebeat.com/2020/12/28/5-ipos-that-show-the-importance-of-data-in-2020/

[8] *Network Monitoring Software*. In: ManageEngine (2021). https://www.manageengine.com/network-monitoring/Eventlog_Tutorial_Part_II.html

[9] Leskiw, A. C.: *Syslog: Servers, Messages & Security– Tutorial & Guide to this System Logs!* In: Network Management Software (2020). https://www.networkmanagementsoftware.com/what-is-syslog/

[10] Al-Fedaghi, S.: *Modeling in Systems Engineering: Conceptual Time Representation*. International Journal of Computer Science and Network Security 21(3), 153–164 (2021)

[11] Bar-Sinai, M., Weiss, G., Marron, A.: *Defining Semantic Variations of Diagrammatic Languages Using Behavioral Programming and Queries*. In: EXE@ MoDELS, pp. 5–11 (2016)

[12] Heidegger, M.: *The Thing*. In: Hofstadter, A. (Trans.) Poetry, Language, Thought, pp. 161–184. Harper and Row (1975)

[13] Santos, O., Kampanakis, P., Woland, A.: *Introduction to and Design of Cisco ASA with FirePOWER Services.* Cisco Press (2016). https://www.ciscopress.com/articles/printerfriendly/2730336

[14] Zafeiroudis, M., Klauzova, V., Gasimov, I.: *Clarify Firepower Threat Defense Access Control Policy Rule Actions.* Cisco (2020). https://www.cisco.com/c/en/us/support/docs/security/firepower-ngfw/212321-clarify-the-firepower-threat-defense-acc.html

[15] Campbell, C., Hoecke, B., Novakovic, D., Acs, G., Duernberger, S.: *Firewall Innovation and Transformation— A Closer Look at ASA and Firepower*. Ciscolive (2017). https://www.ciscolive.com/c/dam/r/ciscolive/emea/docs/2017/pdf/TECSEC-2600.pdf